\documentclass{mic2019}

\usepackage{amssymb}
\usepackage{mathtools}
\usepackage{amsmath}
\usepackage{graphicx}
\usepackage{siunitx}
\usepackage{enumitem}
\usepackage[normalem]{ulem}
\useunder{\uline}{\ul}{}
\usepackage{setspace} % \doublespacing
\usepackage[utf8]{inputenc}
\usepackage{nicefrac}
\usepackage{xspace}
\usepackage{caption}
\usepackage{subcaption}
\usepackage{floatrow}
\usepackage{rotating, lscape, longtable, tabu, booktabs, siunitx, multirow}
\usepackage{cleveref}
\usepackage{verbatim}
\usepackage{nicefrac}
\begin{document}

\title{Algorithms the min-max regret 0-1 Integer Linear Programming Problem with Interval Data}
\author{Iago A. Carvalho\inst{1} \and Thiago F. Noronha\inst{1} \and Christophe Duhamel\inst{2}}
\institute{
    Universidade Federal de Minas Gerais\\
    Department of Computer Science, Belo Horizonte, Brazil\\
    \email{\{iagoac,tfn\}@dcc.ufmg.br}
  \and
    Universit\'{e} Clermont Auvergne\\
    Laboratoire d’Informatique (LIMOS), Clermont-Ferrand, France \\
    \email{christophe.duhamel@isima.fr}
}
\id{id}
\maketitle

\begin{abstract}
We address the Interval Data Min-Max Regret 0-1 Integer Linear Programming problem (MMR-ILP), a variant of the 0-1 Integer Linear Programming problem where the objective function coefficients are uncertain. We solve MMR-ILP using a Benders-like Decomposition Algorithm and two metaheuristics for min-max regret problems with interval data. Computational experiments developed on variations of MIPLIB instances show that the heuristics obtain good results in a reasonable computational time when compared to the Benders-like Decomposition algorithm.
\end{abstract}

\section{Introduction}

The 0-1 Integer Linear Programming problem (ILP, for short) is a well-known NP-Hard mathematical optimization problem, with linear objective function and constraints, in which the domain of all variables is $\{0, 1\}$. An ILP can be formulated by the objective function~\eqref{eq:binaryObj} and the constraints~\eqref{eq:binaryConstraints} and \eqref{eq:binaryDomain}, where $b$ and $c$ are $n$-dimensional vectors of coefficients, $A$ is a $m \times n$-dimensional matrix of coefficients, and $x$ is a $n$-dimensional vector of binary variables. 

\begin{equation} \label{eq:binaryObj}
(\text{ILP}) \quad \min c^T x
\end{equation}
\begin{equation} \label{eq:binaryConstraints}
\qquad Ax \leq b
\end{equation}
\begin{equation} \label{eq:binaryDomain}
\qquad x \in \{0, 1\}^n
\end{equation}

This abstract deals with problems where the coefficients in $c$ are uncertain. We deal with the Interval Data Min-Max Regret 0-1 Integer Linear Programming problem (MMR-ILP, for short). In this problem, the value of the coefficient $c_i$, for all $i \in \{1, \ldots, n\}$, is unknown. However, it is assumed that the value of $c_i$ is in the range $[l_i, u_i]$. A \textit{scenario} is defined as a vector $S = (c^S_1, \ldots, c^S_n)$, where $c^S_i$ is any real value in the interval $[l_i, u_i]$, i.e. a scenario corresponds to a valid assignment of values to the coefficients of variables $x$. There are infinitely many scenarios and MMR-ILP aims at finding a solution that is robust to all of them. MMR-ILP is also NP-Hard optimization problem, since the min-max regret version any problem has, at least, the same complexity of its deterministic counterpart~\cite{Aissi2009}.

MMR-ILP can be formally defined as follows. Let $\Gamma$ be the set of all scenarios, and $\Phi$ be the set of all feasible solutions to the constraints in~\eqref{eq:binaryConstraints} and~\eqref{eq:binaryDomain}. The \textit{regret} of a solution $x \in \Phi$ in a scenario $S \in \Gamma$ is the difference between the cost of $x$ in the scenario $S$ and the cost of the optimal solution $y^S$ for the scenario $S$, i.e. it is the loss of using $x$ instead of $y^S$ if the scenario $S$ occurs. The cost of $x$ in $S$ is denoted by $F(x,S) = \sum_{i = 1}^{n} c_i^S x_i$, while the cost of $y^S$ is denoted by 
$$
   F(S) = \min\limits_{y \in \Phi} F(y,S) = \min\limits_{y \in \Phi} \sum_{i = 1}^{n} c_i^S y_i.
$$ 
The \textit{robustness cost} $Z(x)$ of a solution $x \in \Phi$ is defined as the maximum possible regret of $x$ among all scenarios in $\Gamma$, i.e. 
$$
   Z(x) = \max\limits_{S \in \Gamma} \left\{F(x,S) - F(S)\right\}.
$$
Despite the fact that $|\Gamma|=\infty$, the scenario where the regret of $x$ is the maximum is the scenario $S^{x}$, such that $c^{S^{x}}_i = l_i + (u_i - l_i)x_i$, i.e. $c^{S^{x}}_i = u_i$ if $x_i = 1$, and $c^{S^{x}}_i = l_i$ otherwise~\cite{Aissi2009}. From this result, we have that
$$
  F(x, S^{x}) = \sum_{i = 1}^n u_i x_i, 
$$
$$
  F(y^{S^x},S^{x}) = \min_{y \in \Phi} \sum_{i = 1}^n \big( l_i + (u_i - l_i) x_i \big) y_i, 
$$
and
$$
    Z(x) = F(x, S^{x}) - F(y^{S^x}, S^{x}).
$$
It is worth noticing that $F(y^{S^x},S^{x})$ is still an ILP as in this case $x_i$ is constant. Therefore, the robust cost of a solution $x$ can be computed by solving a single ILP problem in the scenario $S^x$. MMR-ILP aims at finding the solution with minimum robustness cost, i.e. 
$$
   \min_{x \in \Phi} Z(x) = \min_{x \in \Phi} \left\{ F(x, S^{x}) - F(y^{S^x}, S^{x}) \right\}.
$$
 
For the general case, an mathematical formulation for MMR-ILP can obtained by replacing $F(y^{S^{x}}, S^{x})$ with a free variable $\theta$ and adding a new set of linear constrains that bounds the value of $\theta$ to the value of $F(y^{S^{x}}, S^{x})$. The resulting formulation \eqref{eq:nip_obj}-\eqref{eq:nip_domain} has an exponentially large number of constraints. 

\begin{equation} \label{eq:nip_obj}
\min_{x \in \Phi} \sum_{i = 1}^n u_i x_i - \theta
\end{equation}
\begin{equation} \label{eq:nip_constraints}
\qquad \theta \leqslant \sum_{i = 1}^n \left(l_i + (u_i - l_i)x_i\right)y_i, \quad \forall~y \in \Phi
\end{equation}
\begin{equation} \label{eq:nip_domain}
\qquad \theta \in \mathbb{R}        
\end{equation}

\section{Heuristics for the MMR-ILP}

We solved the MMR-ILP using two metaheuristics for interval data min-max regret optimization problems: $(i)$ the Algorithm Mean Upper (AMU)~\cite{Kasperski2006}; and $(ii)$ the Scenario-Based Algorithm (SBA)~\cite{Coco2015,Carvalho2016}. They are described as follows.

\subsection{Algorithm Mean Upper}

AMU is a $2$-approximative heuristic for interval data min-max regret optimization problems. It solves the MMR-ILP into two specific scenarios: the \emph{mean scenario} $s^m$, where the cost of each uncertain coefficient is set to its mean value, \textit{i.e.} $c_{ij}^{s^m} = \frac{l_{ij} + u_{ij}}{2}$, and the \emph{upper scenario} $s^u$, where the cost of each uncertain coefficient is set to its upper value, \textit{i.e.} $c_{ij}^{s^u} = u_{ij}$. AMU computes the robustness cost of the computed solution in each scenario and returns the one which wields the smallest value. 

\subsection{Scenario-Based Algorithm}

SBA is an extension of AMU which inspects a larger number of scenarios. Target scenarios between the lower scenario $s^l \in \Gamma$ (a scenario where the cost of the arcs are set to their respective lower, \textit{i.e.} $c_{ij}^{s^l} = l_{ij}$) and the upper scenario ($s^u$) are investigated. SBA relies on three parameters: the initial scenario $\alpha$; the final scenario $\beta$; and the step size $\gamma$. All parameters are real-valued in the interval $[0, 1]$. Target scenarios are computed as $\alpha + \delta \gamma$, for all $\delta \in \{0, \ldots, i\}$ such that $\alpha + \delta \gamma \leq \beta$. Thus, SBA investigates $\frac{\beta - \alpha}{\gamma}$ different scenarios. One can see that both the mean scenario $s^m$ and the upper scenario $s^u$ are considered by SBA. Thus, SBA produces solutions at least as good as AMU and also holds an approximation ratio of at most 2 for MMR-ILP. The SBA for MMR-ILP uses the parameter settings recommended in Coco et al.~\cite{Coco2015}, being $\alpha = 0.5, \beta = 1.0,$ and $\gamma = 0.05$. Therefore, it inspects a total of 11 scenarios.

\section{Computational experiments}

Computational experiments were carried out on a single core of an Intel Xeon CPU E5645 with $2.4$ GHz clock and $32$ GB of RAM, running under the operating system Linux Ubuntu. ILOG CPLEX solver version $12.6$ is used with default parameters. We employed variations of the classic MIPLIB instances, being the coefficients interval generated as by Carvalho et al~\cite{Carvalho2016}.We assess the quality of AMU and SBA by comparing their results with the primal bound given by the Benders-like Decomposition Algorithm (BDA)~\cite{Montemanni2005}, one of the most successfully exact algorithms for interval data min-max regret optimization problems. We limited the running time of all algorithms to 7200 seconds.

Table~\ref{table:results} show the results of this experiment. The first column reports the BLD average running time and it's standard deviation. The second column shows the AMU average relative deviation regarding the BLD upper bound, computed as $\frac{AMU-BLD}{BLD}$. It also shows the standard deviation deviation of this same metric. The third column presents the AMU average running time and its standard deviation. We show the same information for SBA on the remaining columns.

\begin{table}[h!]
\centering
\caption{Results for BLD, AMU, and SBA when solving the proposed MMR-ILP instances}
\label{table:results}
\begin{tabular}{ccccc}
\toprule
BLD             & \multicolumn{2}{c}{AMU}           & \multicolumn{2}{c}{SBA}           \\ \cmidrule(lr){1-1} \cmidrule(lr){2-3} \cmidrule(lr){4-5}
time (s)        & dev (\%)          & time (s)      & dev (\%)        & time (s)        \vspace{0.2cm} \\ 
4709 $\pm$ 4066 & 10.39 $\pm$ 22.16 & 201 $\pm$ 290 & 9.00 $\pm$ 0.20 & 1983 $\pm$ 2342 \\ \bottomrule
\end{tabular}
\end{table}

One can see from Table~\ref{table:results} that BLD takes an average running time of almost 5000 seconds to run. AMU and SBA relative deviations are very close to each other. However, SBA running time is nearly ten times greater than of AMU. 

We performed a Wilcoxon Sign-Rank Test to verify if there is a significant difference between AMU and SBA relative deviations. The Wilcoxon Test showed that both results do not significantly differ from each other ($p > 0.05$). Therefore, we can conclude that AMU performs better than SBA when solving the proposed MMR-ILP instances since it has a smaller average running time and their relative deviation do not significantly differ.

% \subsection{Abstract}

% The word \textbf{Abstract}, typeset in 10pt bold font, must be placed 0.5cm below the institutions information and an additional space of 0.2cm must be left after it.

% The abstract content must be typeset in 10pt font, and indented by 1.0cm on both the right- and the left-hand sides. The first line of all paragraphs must be additionally indented on the left-hand side by 0.6cm.

% \subsection{References}

% The title References must be typeset according to the subsection rules (except for the indentation) and followed by the list of references. Each reference must be identified by an Arabic numeral in squared parentheses. The list must be ordered alphabetically by the first author.
% The reference text must be indented by 1.0cm with respect to the left margin (excluding the reference number between parentheses).

% \begin{table}
% 	\begin{center}
% 		\begin{tabular}{l r}
% 			\textbf{Algorithm} & \textbf{Average cost} \\
% 			\hline
% 			A & \textbf{1} \\
% 			B & 2 \\
% 			C & 5 \\
% 			\hline
% 		\end{tabular}
% 	\end{center}
% 	\caption{An example table}
% 	\label{tab:example}
% \end{table}

% \begin{figure}
% 	\begin{center}
% 		\includegraphics[width=0.50\textwidth]{mic2019.eps}
% 	\end{center}
% 	\caption{MIC logo}
% 	\label{fig:MIC}
% \end{figure}

\bibliography{example}
\bibliographystyle{plain}

\end{document}